
\font\twelvebf=cmbx10 scaled \magstep1
\hsize 15.2true cm
\vsize 22.0true cm
\voffset 1.5cm
\nopagenumbers
\headline={\ifnum \pageno=1 \hfil \else\hss\tenrm\folio\hss\fi}
\pageno=1

\hfill IUHET--321

\hfill IU/NTC 95--11

\hfill DFTT--67/95

\hfill Oct. 26, 1995
\bigskip

\centerline{\twelvebf Mass sum rules for
singly and doubly heavy-flavored hadrons}

\medskip
\bigskip

\centerline{D. B. Lichtenberg and  R. Roncaglia}

\centerline{Physics Department, Indiana University,
Bloomington, IN 47405, USA}
\centerline{E. Predazzi}

\centerline{Dipartimento di Fisica Teorica,
Universit\`a di Torino and}
\centerline{Istituto Nazionale di Fisica Nucleare,
Sezione di Torino, I-10125, Torino, Italy}

 \vskip 1 cm

Regularities in the hadron interaction energies are
used to obtain formulas
relating the masses of ground-state hadrons,
most of which contain
heavy quarks.  Inputs are the constituent
quark model, the Feynman-Hellmann theorem, and the
structure of the colormagnetic interaction of QCD.
Some of the formulas can also be obtained from
heavy quark effective theory or from diquark-antiquark
supersymmetry.  It is argued
that the sum rules are more general than the
model from which they are obtained.
Where data exist, the formulas agree quite well
with experiment, but
most of the sum rules proposed provide predictions of heavy
baryon masses that will be useful for future measurements.


\bigskip
\medskip \bigskip \bigskip
\centerline{I. INTRODUCTION}
\bigskip

Quantum chromodynamics (QCD)
is the generally accepted theory
of strong interactions, and so, in principle, should be
used to calculate hadron masses. However, thus far, nobody
has succeeded in solving continuum QCD in the
nonperturbative regime necessary to evaluate the
masses of hadrons. This being the case, hadron masses
have been calculated in lattice approximations to
QCD and in various models such as potential and bag
models. These approximations all have their shortcomings,
the lattice calculations requiring a large amount of
computer time to obtain limited results, and the
model calculations being tests of the models rather
than of QCD.

In the light of these difficulties, we  here adopt
an alternative approach of exploiting observed
regularities in the properties of ground-state hadrons
to predictict the masses of hadrons yet to be discovered.
Our procedures should be considered complementary
to other methods used for evaluating hadron masses.
The observed regularities in hadrons are not in the
hadron masses themselves but in
interaction energies obtained by subtracting
the constituent quark masses from the hadron masses.

In particular, it has been
found [1,2] that if reasonable values of the  quark  masses
are subtracted from the masses of the ground-state
vector mesons, then the interaction energy
is a smooth, monotonically decreasing function of the
reduced mass of the constituent quarks. An analogous
result has been found [2,3] for the ground-state baryons
of spin 3/2, where the reduced mass is replaced by
a generalized reduced mass whose inverse is the sum
of the inverse quark masses. These results can be
understood [2,3] from application [4]
of the Feynman-Hellmann theorem [5,6] to the bound-state
systems. A discussion of what constitutes a reasonable
constituent quark mass is given in Ref.\ [2].

It is the purpose of this paper to exploit the regularities
in the hadron interaction energies to obtain sum rules
which relate the masses of different ground-state
hadrons. The advantage in using sum rules
as compared with the treatments in Refs.\ [1--3] is
that the sum rules contain differences of hadron masses
such that the quark constituent masses cancel.
Our sum rules are of three kinds:
those that involve only mesons, those that involve
only baryons, and those that relate the masses of baryons to
the masses of mesons.

Many previous authors have
obtained sum rules relating hadron masses, but, to
our knowledge, most of the ones we present here are
new. As an
early example of a work on heavy baryon sum rules, see
Franklin [7]. A few authors [8,9] have also obtained
sum rules relating mesons and baryons, and
Ref. [9] gives a small subset of the sum rules we obtain
here. Nearly all the formulas  we give involve
one or more hadrons containing heavy quarks.

We motivate our sum rules from the constituent
quark model, but
they appear to hold because of regularities in
hadron masses as a function of their valence quark
content. Furthermore, some of the formulas we obtain also
follow from heavy quark effective theory [10,11]
or from approximate antiquark-diquark supersymmetry [12],
sometimes called superflavor symmetry [13].
Therefore, we believe most of our
sum rules are more general than
their derivation would suggest. It must be stressed,
in fact, that our results do not follow
from any specific model (such as potential, lattice, flux
tube etc.) but, as already mentioned, from an
extended survey of the regularities of the hadron
spectrum in flavor space and from general theorems.
This technique may not be the most satisfactory one from the
mathematical point of view (since our sum rules are not,
strictly, derived from first principles),
but, on the other hand, as emphasized by
Martin and Richard
in the context of one of our earlier predictions
(obtained in similar way), our result
``...not surprisingly, comes very
close to the (preliminary) experimental mass'' [14].
The effectiveness of our sum rules is,
we believe, a sufficient motivation
to propose them. They will prove
extremely useful when the systematic search for heavy
hadrons will have reached maturity: it will then be very
convenient to have a consistent set of predictions not
linked to any specific
theoretical model but based rather on general properties.

As we have remarked, application of
the Feynman-Hellmann theorem provides motivation for
the systematic study of the vector mesons and spin-3/2
baryons. However, experimental regularities also appear
in the masses of pseudoscalar mesons and spin-1/2 baryons.
For this reason we also propose sum rules involving
the masses of these particles. Still other sum rules
concern the masses of
ground-state hadrons averaged over their various
spin states. For most mesons and for those
 baryons containing three different quark flavors, we
can obtain unique expressions for the spin-averaged masses
in terms of the physical hadron  masses [15].
Our sum rules for spin-averaged masses are restricted
to those cases in which spin averaging is possible
in terms of physical masses.

In order to obtain spin averages, we assume, first, that
the spin splittings of hadrons with a given quark
content arise from the colormagnetic interaction
of QCD; and, second, that the colormagnetic interaction
can be treated perturbatively [16]. The first of these
assumptions is good for the ground-state
hadrons because, in first approximation,
the hadron wave functions  have
no orbital angular momentum, and so tensor and spin-orbit
forces do not contribute. Because the colormagnetic
interaction between two quarks goes inversely as the
product of their masses, the perturbative approximation
improves as the quark masses increase. However, even
though the  perturbative approximation might not be good
for hadrons containing only light quarks, nevertheless,
the empirical
regularities in hadron interaction energies persist.
Because our mass formulas really rely on these regularities
and not  on perturbation theory, these formulas should
be good even where a perturbative treatment
of the colormagnetic interaction breaks down.

We can somewhat improve the perturbative
expression of QCD for the colormagnetic splitting between
different hadron spin states. We accomplish this
by including  parameters whose values are determined
by fits to known data, obtaining semiempirical
mass formulas for meson and baryons spin splittings
[17,2,3]. Using
these formulas, we can, for example, predict the masses
of all hadrons of a given spin multiplet if the mass
of one member is known from experiment. These results
give us additional guidance in obtaining sum rules.

In Sec.\ 2 we discuss our methods in more detail.
In Sec.\ 3 we give sum rules for
mesons, in Sec.\ 4 we give sum rules for baryons, and
in Sec.\ 5 we give sum rules relating the masses
of baryons to those of mesons. In Sec.\ 6 we
test a few of our sum rules against known data
[18,19] and use other formulas to give
predictions for the masses of as yet unobserved
hadrons. We also discuss our results.

\bigskip
\centerline{II. OBTAINING HADRON MASS SUM RULES}
\bigskip

The mass $M_M$ of a meson and the mass $M_B$ of a baryon
presumably can be calculated in principle from QCD
and can be approximately calculated in practice from
lattice QCD or from a potential model or other
model motivated by QCD. However, here we take a
different approach, and exploit regularities in the
masses of ground-state hadrons [2,3] to obtain
sum rules relating their masses.

According to QCD, a meson contains a valence quark and
antiquark plus a sea of gluons and quark-antiquark pairs.
Similarly, a baryon contains three valence quarks plus
a sea of gluons and pairs.
In the constituent quark model, we neglect the
sea, and take a hadron to be composed only of valence
quarks with constituent masses that are a few hundred
MeV larger than current quark masses.
We assume that the constituent quarks can be assigned
unique masses independent of the hadrons in which
the quarks are bound. As we shall see, this turns out to
be a very fruitful
and economical assumption even though, strictly,
it is probably only approximately true.

Within this framework, we
introduce meson and
baryon {\it interaction energies} $E(ij)$ and $E(ijk)$,
defined by
$$E(ij)= M_M(ij) -m_i -m_j, \eqno(1)$$
$$E(ijk)= M_B(ijk) - m_i - m_j -m_k,  \eqno(2)$$
where the symbols $i$, $j$, and $k$ denote the valence
quarks (or antiquarks) and $m_i$, $m_j$, and $m_k$ denote
their constituent masses. It has been
shown [2, 3] that, for
``reasonable'' choices of the
constituent quark masses, the
interaction energies $E(ij)$ of observed vector mesons
and the $E(ijk)$ of observed spin-3/2 baryons are smooth,
monotonically decreasing functions of a generalized
reduced mass $\mu$:
$$dE(ij)/d\mu \leq 0, \quad dE(ijk)/d\mu \leq 0, \eqno(3)$$
where $\mu$ is defined by
$$1/\mu = \sum_i 1/m_i. \eqno(4)$$
The fits to the data in Refs.\ [2] and [3] also have
the property that the interaction energies are
concave upward as a function of $\mu$:
$$d^2E(ij)/d\mu^2 \geq 0, \quad d^2E(ijk)/d\mu^2 \geq 0.
\eqno(5)$$
Furthermore, it was shown in Ref. [3] that the
inequalities \ (3) and (5)
still hold if the constituent quark masses are taken
to be the same in mesons and baryons. The regularities
in the interaction energies can be generalized to
as yet undiscovered mesons and baryons.

The motivation for examining the dependence of
the interaction energies as a function of $\mu$
comes from the Feynman-Hellmann theorem [5,6]. This
theorem enables us to obtain the inequalities (3)
in a Hamiltonian formalism with certain restrictions
on the flavor-dependence of the interaction [2,3].

It was perhaps not emphasized sufficiently
in Refs. [2] and [3]
that {\it experimental} values [18,19] of the $M_M(ij)$
and $M_B(ijk)$ lead to the inequalities (3) and (5)
provided the quarks are assigned
constituent mass values which satisfy certain
constraints obtained from the Feynman-Hellmann theorem.
But the experimental hadron masses
presumably are calculable in full QCD and do not
depend on the constituent quark model. Therefore,
the smoothness of the interaction energies $E(ij)$ and
$E(ijk)$ defined in Eqs. (1) and (2)
cannot really
depend on the constituent quark model, even though
constituent quark masses appear in these equations.
Experiment is telling us simply that if reasonable
values of constituent quark masses are subtracted
from ground-state vector meson or spin-3/2 baryon
masses, the resulting quantities are smooth, monotonically
decreasing functions of $\mu$.
The use of the phrase ``...reasonable
values of constituent quark masses" hides a significant
subtlety. It turns out, in fact, that fits  to the
spectroscopic data do not well determine
{\it absolute} values of the constituent quark masses [2].
Much more constrained
from the data are the quark mass {\it differences}.
Nevertheless, as shown in Ref.\ [2] one finds that
some sets of quark masses found in the literature are
compatible with our general theoretical considerations
while other sets are not.

The principal contribution to the spin-dependent splitting
of ground-state hadrons with the same quark content
comes from the colormagnetic interaction of QCD [16].
In Refs.\ [2] and [3] it was shown that the structure
of the colormagnetic interaction is such that
the Feynman-Hellmann theorem does not require that
that the interaction energies $E(ij)$ and $E(ijk)$ of the
pseudoscalar mesons  and spin-1/2 baryons be
monotonically decreasing functions of $\mu$.
Nevertheless, we find empirically that even in these
cases some regularities exist in the hadron interaction
energies. This fact allows us to propose
additional sum rules for pseudoscalar mesons and spin-1/2
baryons. Furthermore, we propose still other sum rules
for hadron masses averaged over spin, some of which
have appeared previously  [9].
Because of the structure of
the colormagnetic interaction, the spin averages of
mesons and of baryons containing quarks of three different
flavors can be taken with a
unique result in a perturbative approximation [15].

We also make use of a result of Bertlmann and Martin [20]
that in a  nonrelativistic approximation, the spin-averaged
meson masses satisfy the inequalities ($m_i< m_j < m_k$):
$$2M(ij) - M(ii) - M(jj) \geq 0, \eqno(6)$$
$$M(ii) + M(jk) - M(ij) -M(ik) \leq 0. \eqno(7)$$
Furthermore, we use a generalization of (6) and
(7) to baryons [21]:
$$2 M(ijk) - M(ijj) - M(ikk) \geq 0. \eqno(8)$$
$$M(iii) + M(ijk) - M(iij) - M(iik) \leq 0. \eqno(9)$$
Where data are available, it turns out that the
inequalities (6)--(9) hold also for hadrons of
definite spin configuration. We conjecture that these
inequalities will hold for as yet undiscovered ground-state
hadrons. Using (1) and (2), we see that
(6)--(9) should hold for the corresponding
meson and baryon interaction energies because the quark
masses cancel in these expressions
(which, among other things, greatly reduces the ambiguities
that could follow from the aforementioned freedom in choosing
an {\it absolute} scale of quark masses).

We use these ideas to motivate our proposed sum
rules for hadron masses. In particular,
we use quantitative estimates of the behavior of
$E(ij)$, $E(ijk)$ and their first and second derivatives
taken from the known hadrons and extrapolate to unknown
hadrons.  All the sum rules of this
paper contain the masses of three or four hadrons
except those involving spin-averaged masses.
In order to obtain spin averaged masses, we must double
the number of mesons and triple
the number of baryons appearing in each sum rule [15].

Using the methods of this paper we can
obtain still more complicated sum rules involving
even more hadrons, but believe it is  not
interesting to exhibit them here.

\bigskip
\centerline{III. MESON SUM RULES}

\bigskip
We begin with the mesons, for which we can find only
a few sum rules. From the Feynman-Hellmann theorem
we deduce that inequality (6) is most likely to
be an equality in the case in which the value of
$\mu$ in the different terms varies least. This is
the case in which the $i$ quark is either $u$ or $d$
and the $j$ quark is $s$. (We neglect the mass difference
between $u$ and $d$, and denote these quarks by $q$.)
Using (6) in the form of an equality for vector
mesons, we obtain
$$2K^*(qs) -\phi(ss)- \rho(qq) = 0, \eqno(10)$$
where here and subsequently we
let the symbol
for a hadron denote its mass. We put the
quark content in parentheses only the first time
we use the symbol for a hadron,  and we omit
the bar on the symbol for an antiquark.

This sum rule
agrees quite well with the data [18], but
it does not involve mesons containing heavy quarks
and may not be new.
Because of the larger changes in $\mu$ for
the heavy $c$ or $b$ quarks, sum rules analogous to
Eq.\ (9) do not hold for mesons containing these quarks,
but only the inequality (6).
In obtaining (10), we assume
that the $\phi$ is an $s\bar s$ state. We cannot obtain
a corresponding sum rule for the pseudoscalars, because
neither the $\eta$ nor the $\eta'$ is a pure $s\bar s$
state. (They are mixtures of $q\bar q$ and $s\bar s$
and may contain some admixture of glueball as well.)

A sum rule
(connected with Eq.\ (7)) for vector mesons containing one
heavy quark, is
$$D_s^*(sc) - D^*(qc) + B^*(qb) - B_s^*(sb) = 0
\eqno(11.1)$$
An analogous formula for
pseudoscalar mesons with the same quark content is
$$D_s - D + B - B_s = 0. \eqno(11.2)$$

In our way of reasoning,
these sum rules hold approximately because
the values of $\mu$ for the different mesons are such
that the interaction energies cancel to a good
approximation. We are led to propose them by examining
the systematics of the variation of the interaction
energy with $\mu$.
According to heavy quark effective theory [10,11,13],
the interaction energy in a hadron
should not change appreciably
when a $c$ quark is replaced by a $b$ quark. Because
the quark masses cancel in (11), this
sum rule follows immediately from heavy quark theory
for the spin-averaged masses. We are proposing the
stronger condition that analogous sum rules hold separately
for the vectors and pseudoscalars. The separate sum
rules follow from heavy quark effective theory if
the difference in  the colormagnetic energy
between a meson containing a $c$ quark and a meson
containing a $b$ quark is neglected.

All the particles appearing in the combinations
(11.1,2) have by now been identified experimentally
so that Eqs.\ (11.1,2) can be tested. The result of
the comparison with the data is shown in Table I (where all
the comparisons between the data, when available, and the
sum rules to be discussed in what follows, are also given).

\bigskip
\centerline{IV. BARYON SUM RULES}

\bigskip
We have many more possibilities to obtain sum rules for
baryons than for mesons
because for a fairly large number of baryons,
the variation of the generalized $\mu$ is small.
Examining the systematics, we find that (8) is
approximately an equality if none of the quarks is
heavy. We obtain the following two sum rules for
spin-3/2 baryons (as already noted, we put the
quark content in parentheses only the first time
the symbol for a hadron appears in this paper):
$$2 \Sigma^*(qqs) - \Delta(qqq) - \Xi^*(qss) = 0,
\eqno(12)$$
$$2 \Xi^* - \Sigma^* - \Omega(sss) = 0,
\eqno(13)$$
These two sum rules are well known,
and together are just the
Gell-Mann--Okubo baryon decuplet mass formula.

We next turn to sum rules involving at least two
baryons  containing  a heavy quark.
We obtain the following sum rules from the systematics:
$$\Omega + \Sigma_b^*(qqb) - \Xi^* -\Xi_b^*(qsb) = 0,\eqno(14.1) $$
$$\Sigma_c^*(qqc) + \Omega_b^*(ssb) - \Xi_c^*(qsc)
- \Xi_b^* =0,                                        \eqno(14.2) $$
$$\Xi_c^*+\Xi_b^*-\Omega_c^*(ssc)-\Sigma_b^* = 0,    \eqno(14.3) $$
$$\Omega + \Xi_c^* - \Xi^* - \Omega_c^*= 0,          \eqno(14.4) $$
$$\Xi^* + \Sigma_b^* - \Sigma^* - \Xi_b^* = 0,       \eqno(14.5) $$
$$ \Xi^* + \Xi_c^* - \Sigma^* - \Omega_c^* = 0.      \eqno(14.6) $$
Because each of these sum rules contains the mass
of a baryon not yet discovered, they cannot be tested
at this time. On the other hand, these
relations (like most of those that follow) provide an
approximate value for the masses of the as yet unknown
baryons and will therefore await experimental verification.
Our predictions are summarized in Table II.

We also have sum rules involving spin-1/2 baryons.
They are somewhat different from those in Eq.\ (14),
however, because the  colormagnetic energies, which
depend on quark masses and spin configuration,
are different. If all three quarks in a baryon
have different flavors, then two distinct baryons
exist with a given quark content.
In order to
distinguish between these two states, it is convenient
to order the quarks in a baryon such that the two
lightest quarks are the first two.
For example, in the
case of the $\Lambda$ and $\Sigma^0$, the first two quarks
are $u,d$ and the third quark is $s$.
In the $\Lambda$, the first two quarks
have spin 0, and in the $\Sigma$, the first two
quarks have spin 1. Because we do not distinguish
between the mass of the $u$ and $d$ quark, we write
the quark content  of the $\Lambda$ and $\Sigma$
as $qqs$. For any pair of spin-1/2 baryons containing
three different flavors of quarks, the lighter
baryon is the one
in which  the first two quarks have spin 0, and the
heavier baryon is the one in which the first two
quarks have spin 1,
as with the $\Lambda$ and $\Sigma$.
We have not been able to find any sum rules involving
four baryons in the case in which the first
two quarks have spin 0 ($\Lambda$-type symmetry).

The sum rules for the case in which the first two
quarks have spin 1 ($\Sigma$-type symmetry) are
$$\Sigma(qqs) + \Omega_c(ssc) - \Xi(ssq) - \Xi_c'(qsc) = 0, \eqno(15.1)$$
$$\Xi + \Sigma_c(qqc) - \Sigma - \Xi_c' = 0,                \eqno(15.2)$$
$$\Xi + \Xi_b'(qsb) - \Sigma - \Omega_b(ssb) = 0,           \eqno(15.3)$$
$$\Xi_c' + \Xi_b' - \Sigma_b(qqb) - \Omega_c = 0,           \eqno(15.4)$$
$$\Omega_c + \Sigma_b - \Sigma_c - \Omega_b = 0.            \eqno(15.5)$$
(Here and in the following, when two spin-1/2 baryons exist
with the same quark content
and the same Greek symbol [18],
a prime denotes the configuration in
which the first two quarks have spin 1.)

Only the last of the sum rules in (15) follows from
heavy quark theory, and then only in the approximation
of neglecting certain differences in colormagnetic
energy. Data do not yet exist to test these sum rules.
We expect that all of the mass formulas in (14) and (15)
will hold to about 20 MeV or better.
Again, when applicable, our predictions are
listed in Table II.

We have been able to find a rather large number of
sum rules involving some baryons
which contain two heavy quarks. Although several
theoretical papers have been written about
baryons containing two heavy quarks
[7,8,13,22,23], none has yet been observed.
We list our  sum rules for these baryons in an Appendix.

\bigskip
\centerline{V. SUM RULES FOR BARYONS AND MESONS}

\bigskip
We have obtained an even larger number of mass formulas
involving two baryons and two mesons than the number
involving four baryons. We give here those in which
no baryon contains more than one heavy quark, and
relegate to the Appendix the formulas involving baryons
containing two heavy quarks.

The formulas are of four kinds: 1) those with spin-3/2
baryons and vector mesons, 2) those with spin-1/2 baryons
($\Sigma$-type symmetry) and pseudoscalar mesons, 3) those
with spin-1/2 baryons ($\Lambda$-type symmetry) and
pseudoscalar mesons, and 4) those involving spin-averaged
baryons and mesons.

We begin with formulas for spin-3/2 baryons and spin-1
mesons. Our sum rules are

$$ \Xi^* -  \Xi_c^* + D^*(qc) - K^* = 0,      \eqno(16.1)$$
$$ \Omega - \Omega_b^* + B^*(qb) - K^* = 0,    \eqno(16.2)$$
$$ \Sigma_c^* - \Xi^* + \phi - D^* = 0,       \eqno(16.3)$$
$$ \Sigma_c^* - \Xi_c^* + K^* - \rho = 0,     \eqno(16.4)$$
$$ \Sigma_c^* - \Sigma_b^* + B^* - D^* = 0,   \eqno(16.5)$$
$$ \Omega_c^* - \Xi^* + \rho - D^* = 0,       \eqno(16.6)$$
$$ \Omega_c^* - \Xi_c^* + K^* - \phi = 0,     \eqno(16.7)$$
$$ \Omega_c^* - \Xi_b^* + B^* - D_s^*(sc) = 0, \eqno(16.8)$$
$$ \Sigma_b^* - \Sigma_c^* + D_s^* - B_s^*(sb) = 0,
\eqno(16.9)$$
$$ \Xi_b^* - \Omega + \phi - B^* = 0,      \eqno(16.10)$$
$$ \Xi_b^* - \Omega_b^* + K^* - \rho = 0,   \eqno(16.11)$$
$$ \Omega_b^* -  \Omega_c^* + D^* - B^* = 0,  \eqno(16.12)$$
$$ \Omega_b^* - \Sigma_b^* + \rho - \phi = 0. \eqno(16.13)$$

Of these sum rules,
(16.5), (16.8), (16.9), (16.10), and (16.12)
can be justified by antiquark-diquark supersymmetry.
If the contribution from heavy quarks to the colormagnetic
energy is neglected, then (16.5), (16.9), and (16.12)
also follow from heavy quark effective theory.
The masses of all hadrons appearing in
the sum rules (16.3,5,9) are known from
experiment, and these sum rules are well satisfied
(as shown in Table I).

The formulas involving two spin-1/2 baryons ($\Sigma$-type
symmetry) and two pseudoscalar mesons are:

 $$ \Xi_c' - \Omega_b + B_s - D = 0,       \eqno(17.1)$$
 $$ \Sigma_b - N(qqq) + K - B_s = 0,       \eqno(17.2)$$
 $$ \Xi_b' - \Sigma_c + D - B_s = 0.       \eqno(17.3)$$

Of these, the first and the third descend from
antiquark-diquark supersymmetry.

We were unable to find formulas involving two spin-1/2 baryons
($\Lambda$-type symmetry) containing only up to one heavy quark,
and two pseudoscalar mesons. The formulas involving baryons
containing two heavy quarks are reported in the appendix.

Finally, the sum rules involving spin-averaged baryons and
mesons are:
$$ (2\Sigma^*+\Sigma+\Lambda(qqs))/4 - (N+\Delta)/2 +
   (3\rho +\pi)/4 - (3K^*+K(qs))/4 = 0,   \eqno(18.1)$$
$$ (2\Sigma_c^*+\Sigma_c+\Lambda_c)/4 -
   (2\Sigma^*+\Sigma+\Lambda)/4 + (3K^*+K)/4-(3D^*+D)/4 = 0,
\eqno(18.2)$$
$$ (2\Sigma_c^*+\Sigma_c+\Lambda_c)/4
 - (2\Sigma_b^*+\Sigma_b+\Lambda_b(qqb))/4 +
   (3B^*+B)/4 - (3D^*+D)/4 = 0, \eqno(18.3)$$
$$ (2\Sigma_b^*+\Sigma_b+\Lambda_b)/4 -
   (2\Sigma_c^*+\Sigma_c+\Lambda_c)/4 + (3D_s^*+D_s/4) -
   (3B_s^*+B_s)/4 = 0.       \eqno(18.4)$$
The first three of these equations have been obtained
previously [9]. The last two can also be derived
from heavy quark theory. Furthermore, Eq. (18.4) follows
from (18.3) with the help of (11.1) and (11.2). We do,
however, have several new sum rules involving spin
averages of baryon and meson masses. These are given
in the Appendix in Eq. (A.6).

Those sum rules for which data exist compare quite well
with experiment, as can be seen from Table I.
We expect that the sum rules of this section will
be good to 20 MeV or better.
\bigskip
\centerline{VI. RESULTS AND DISCUSSION}

\bigskip

We have not verified that all the sum rules of this
paper are  linearly independent, but we believe
that most, if not all, of them are, with the
exception of the formula given in Eq.\ (18.4),
which we have explicitly pointed out is derivable
from others. If a few sum rules should turn out
to be linear combinations of others, no harm is done.

Some of the formulas given in the preceding three
sections contain only masses of known hadrons. We
test these formulas using
data from the Particle Data group [18] and
more recent preliminary data from a
conference talk by Jarry [19]. We give our results
in Table I. As can be seen from this table, those of
our sum rules which can be tested agree with experiment
within about 10 MeV or better.

If all but one of the
hadrons entering  a formula have been observed,
we use the formula to predict the mass of the unknown
hadron.
We go on to use these predicted masses in other sum
rules to obtain still further predictions.
In following
this method, we use the sum rules of the appendix
as well as those appearing in the main body of the paper.
The predicted masses arising from this procedure are
given in Table II. Our estimated errors
are 20 MeV or less for sum rules not involving
any  baryons containing two heavy quarks.
The caption to Table II
gives our estimated errors in all cases.

It is evident from Table II that we are unable to use our
mass formulas to get predictions for the
baryons  $\Xi_{bb}$ and
$\Omega_{bb}$. However, we can estimate the difference
$\Omega_{bb}-\Xi_{bb}$. From heavy quark symmetry, it
should be equal to $\Omega_{cb}'-\Xi_{cb}'= 95$ MeV, while
antiquark-diquark supersymmetry suggests that
$\Omega_{bb}-\Xi_{bb}= B_s - B = 90$ MeV.

We may consider the question: If any of our sum rules
should turn out to be badly in error, would we learn
anything? First, we believe it is highly unlikely that
such an event will happen, because enough hadrons
are already known to give us confidence that the
regularities in the interaction energies are much
more than coincidence. Therefore, these regularities
should persist
in ground-state hadrons not yet discovered. Second, in the
unlikely event that an
exception is found, it will cast doubt on the
flavor-independence of the fundamental interaction
between quarks.

In conclusion, relying on observed systematics of
the interaction energies of known hadrons, we have
obtained a large number of sum rules for the masses
of known and unknown hadrons.  In those
cases in which the masses of all hadrons entering our
sum rules are known from experiment, the sum rules
agree with experiment to about 10 MeV or better. This
fact gives us confidence that the predictions of
unknown hadron masses which follow from the sum rules
are likely to be correct within quite small errors
compared to the masses themselves. We believe our
predictions should be a useful guide to experimentalists
searching for new hadrons, as our results depend
on the regularities in observed hadrons persisting
to hadrons not yet seen rather than
on any specific model of quark interactions.
\bigskip
\centerline{ACKNOWLEDGMENTS}
\medskip
We thank Rick Van Kooten and Saj Alam for information
about the latest experimental data.
Part of this work was done while
one of us (EP)  visited Indiana University. He
thanks the members of the physics department for their
kind hospitality.  This work was supported in part by
the U.S. Department of Energy, in part by the U.S.
National Science Foundation, and in part by the
Italian National Institute for Nuclear Physics and
by the MURST
(Ministry of Universities, Research, Science and Technology) of Italy.

\bigskip
\centerline{APPENDIX}
\bigskip
In this Appendix we give sum rules involving at least
two baryons containing two heavy quarks. We expect
that the sum rules involving baryons containing no
$b$ quarks are good to 30 MeV, those involving baryons
containing no more than one $b$ quark are good to
about 40 MeV, and those involving baryons containing
two $b$ quarks are good to about 50 MeV.

The formulas for spin-3/2 baryons are:
$$ \eqalign{
\Omega + \Xi_{cc}^*(ccq) - \Xi_c^* - \Omega_c^* = 0,           \cr
\Omega + \Xi_{cb}^*(qcb) - \Xi_c^* - \Omega_b^* = 0,           \cr
\Sigma_c^* + \Xi_{cb}^* - \Sigma_b^* - \Xi_{cc}^* = 0,         \cr
\Sigma_c^*+\Omega_{cb}^*(scb)-\Sigma_b^*-\Omega_{cc}^*(ccs)=0, \cr
\Sigma_c^* + \Xi_{bb}^*(bbq) - \Sigma_b^*-\Xi_{cb}^* = 0,      \cr
2 \Xi_c^* - \Xi^* - \Xi_{cc}^* = 0,                    \cr
\Omega_c^* + \Sigma_b^* - \Xi^* - \Xi_{cb}^* = 0,              \cr
\Omega_c^* + \Xi_b^* - \Omega - \Xi_{cb}^* = 0,                \cr
\Sigma_b^*+\Omega_{cb}^*-\Sigma_c^*-\Omega_{bb}^*(bbs)=0,      \cr
\Omega_b^* + \Xi_{cb}^* - \Omega_c^* - \Xi_{bb}^* = 0.         }
\eqno(A.1)$$

The formulas for spin-1/2 baryons ($\Sigma$-type symmetry) are:
$$ \eqalign{
\Xi_c' + \Xi_{cb}'(qcb) - \Xi_b' - \Xi_{cc}(ccq) = 0,         \cr
\Sigma_b+\Omega_{cc}(ccs)-\Sigma_c-\Omega_{cb}'(scb) = 0,     \cr
\Omega_b + \Xi_{cc} - \Xi_b' - \Omega_{cc} = 0,               \cr
\Omega_b + \Omega_{cc} - \Omega_c - \Omega_{cb}' = 0.         }
\eqno(A.2)$$

We now turn to the mass sum rules involving two baryons and
two mesons.  For spin-3/2 baryons and vector mesons, we have:
$$ \eqalign{
\Sigma_c^* - \Xi_{cb}^* + B^* - \rho = 0,           \cr
\Xi_c^* - \Xi_{cc}^* + D^* - K^* = 0,               \cr
\Xi_c^* - \Omega_{cc}^* + D_s^* - K^* = 0,          \cr
\Xi_c^* - \Xi_{cb}^* + B^* - K^* = 0,               \cr
\Xi_c^* - \Omega_{cb}^* + B_s^* - K^* = 0,          \cr
\Omega_c^* - \Xi_{cb}^* + B^* - \phi = 0,           \cr
\Omega_c^* - \Omega_{cb}^* + B_s^* - \phi = 0,      \cr
\Sigma_b^* - \Xi_{cb}^* + D^* - \rho = 0,           \cr
\Sigma_b^* - \Xi_{bb}^* + B^* - \rho = 0,           \cr
\Xi_{cc}^* - \Sigma_c^* + K^* - D_s^* = 0,          \cr
\Omega_{cc}^* - \Xi_{cc}^* + B^* - B_s^* = 0,       \cr
\Omega_{cc}^* - \Omega_{cb}^* + B^* - D^* = 0,      \cr
\Xi_{cb}^* - \Xi_b^* + \phi - D_s^* = 0,            \cr
\Xi_{cb}^* - \Omega_{bb}^* + B_s^* - D^* = 0,       \cr
\Omega_{cb}^* - \Omega_{bb}^* + B^* - D^* = 0,      \cr
\Omega_{bb}^* - \Omega_{cb}^* + D_s^* - B_s^* = 0.  }
\eqno(A.3)$$

The sum rules involving spin-1/2, $\Sigma$-type symmetry
baryons and pseudoscalar mesons are:
$$ \eqalign{
N - \Xi_{cb}' + B_c - \pi = 0,                    \cr
\Sigma - \Xi_{cc} + \eta_c(cc) - K = 0,           \cr
\Omega_{cc} - \Xi_{cc} + D - D_s = 0,             \cr
\Xi_{cb}' - \Omega_{cb}' + B_s - B = 0,           \cr
\Xi_{bb}(bbq) - \Omega_{bb}(bbs) + B_s - B = 0.   }
\eqno(A.4)$$

The sum rules involving spin-1/2, $\Lambda$-type symmetry
baryons and pseudoscalar mesons are:
$$ \eqalign{
\Xi_{cb}(qcb) - \Omega_{cb}(scb) + D_s - D = 0,   \cr
\Omega_{cb} - \Xi_{cb} + B - B_s = 0.             }
\eqno(A.5)$$

The formulas involving spin averages are:
$$ \eqalign{
(2\Sigma_c^*+\Sigma_c+\Lambda_c)/4 -
   (2\Xi_{cb}^*+\Xi_{cb}'+\Xi_{cb})/4  +
   (3B_s^*+B_s)/4 - (3K^*+K)/4 = 0,                  \cr
(2\Sigma_b^*+\Sigma_b+\Lambda_b)/4 -
   (2\Xi_{cb}^*+\Xi_{cb}'+\Xi_{cb})/4  +
   (3D_s^*+D_s)/4 - (3K^*+K)/4 = 0,                  \cr
(2\Xi_{cb}^*+\Xi_{cb}'+\Xi_{cb})/4  -
   (2\Omega_{cb}^*+\Omega_{cb}'+\Omega_{cb})/4 +
   (3D_s^*+D_s)/4 - (3D^*+D)/4 = 0,                  \cr
(2\Xi_{cb}^*+\Xi_{cb}'+\Xi_{cb})/4  -
   (2\Omega_{cb}^*+\Omega_{cb}'+\Omega_{cb})/4 +
   (3B_s^*+B_s)/4 - (3B^*+B)/4 = 0.                  }
\eqno(A.6)$$

\vfill\eject

References
\bigskip

\item{[1]} W. Kwong and J. L. Rosner, Phys.\ Rev.\ D
{\bf 44}, 212 (1991).

\item{[2]} R. Roncaglia, A. Dzierba, D.B. Lichtenberg,
and E. Predazzi, Phys. Rev. D {\bf 51}, 1248 (1995).

\item{[3]} R. Roncaglia, D.B. Lichtenberg,
and E. Predazzi, Phys.\ Rev.\ D {\bf 52},  1722 (1995).

\item{[4]} C. Quigg and J. L. Rosner, Phys. Rep. {\bf 56}
(1979) 167.

\item{[5]} R.~P. Feynman, Phys. Rev. {\bf 56}, 340 (1939).

\item{[6]} H. Hellmann, Acta Physicochimica URSS
I, 6, 913 (1935); IV, 2, 225 (1936);
Einf\"uhrung in die Quantenchemie (F. Deuticke,
Leipzig and Vienna, 1937) p. 286.

\item{[7]} J. Franklin, Phys. Rev. D {\bf 12}, 2077 (1975)
and references therein.

\item{[8]} T. Ito, T. Morii, M. Tanimoto, Z. Phys. C
{\bf 59}, 57 (1993).

\item{[9]} D.B. Lichtenberg and R. Roncaglia,
Phys.\ Lett.\ B {\bf 358}, 106 (1995).

\item{[10]} N. Isgur and M. B. Wise, Phys. Lett.\
B {\bf 232}, 113 (1989); {\bf 237}, 527 (1990).

\item{[11]} E. Eichten, B. Hill, Phys.\ Lett.\ B
{\bf 234}, 511 (1990);
B. Grinstein, Nucl.\ Phys.\  {\bf B339}, 253 (1990);
H. Georgi, Phys.\ Lett.\ B {\bf 240}, 447 (1990).

\item{[12]} H. Miyazawa, 1966, Prog.\ Theor.\
Phys. {\bf 36}, 1266 (1966);
D.B. Lichtenberg, J. Phys.\ G {\bf 16}, 1599 (1990);
{\bf 19}, 1257 (1993).

\item{[13]} H. Georgi and M.B. Wise, Phys.\ Lett.\ B
{\bf 243}, 279 (1990); M.J. Savage and M.B. Wise,
Phys.\ Lett.\  B {\bf 248}, 177 (1990).

\item{[14]} A. Martin and J.-M. Richard, Phys.\ Lett.\ B
{\bf 355}, 345 (1995).

\item{[15]} M. Anselmino, D. B. Lichtenberg, and E.
Predazzi, Z. Phys.\ C {\bf 48}, 605 (1990).

\item{[16]} A. De R\'ujula, H. Georgi, and S. L. Glashow,
Phys.\ Rev.\ D {\bf 12}, 147 (1975).

\item{[17]} X. Song, Phys.\ Rev.\ D {\bf 40}, 3655 (1989);
Yong Wang and D. B. Lichtenberg,
Phys.\  Rev.\ D {\bf 42}, 2404 (1990).

\item{[18]} Particle Data Group: L. Montanet et al.,
Phys.\ Rev.\ D {\bf 50}, 1173 (1994).

\item{[19]} P. Jarry,  XV International Conference
on Physics in Collision, Cracow, Poland (June 8--10,
1995).

\item{[20]} R.A. Bertlmann and A. Martin, Nucl. Phys.
 {\bf B168}, 111 (1980).

\item{[21]} E. Bagan, H.G. Dosch, P. Gosdzinsky,
S. Narison, J.-M. Richard, Z. Phys. C {\bf 64} (1994) 57.

\item{[22]} J.-M. Richard, in {\it The future of
High Sensitivity Charm Experiments}, Proc. of
the Charm2000 Workshop, Batavia (June, 1994), edited
by D. M. Kaplan and S. Kwan, Batavia, Ill. (1994) p. 95.

\item{[23]} M.L. Stong, U. of Karlsruhe Report TTP 95-02
(1995).

\item{[24]} M. S. Alam, talk at {\it Baryons '95},
Santa Fe, New Mexico (October, 1995) and private
communication.
\vfill \eject

TABLE I.~~~Test of sum rules when all particles' masses are
known from experiment.  Column 1 refers to the
equation number; column 2 lists the actual value in MeV
of the violation of the mass formula and its error,
obtained by using experimental mass values from
the Particle Data Group [18] and from a recent conference
report by Jarry [19].

\vskip .3cm
$$\vbox {\tabskip=0pt \offinterlineskip
\halign {\strut\tabskip=1em plus2em\hfil #\hfil\quad &
&\quad\hfil # \hfil\quad  \cr
\noalign{\hrule}
\noalign{\hrule}
           &                     \cr
  Eq.\ \#  & \hfil Violation in MeV\hfil \cr
           &              \cr
\noalign{\hrule}
           &              \cr
   (10)    &  $~1 \pm ~3$ \cr
  (11.1) & $5\pm ~5$   \quad {\rm or} \quad
$12 \pm ~4^*$ \cr
  (11.2) & $4\pm ~5$  \quad  {\rm or} \quad
$11 \pm ~2^*$ \cr
   (12)    &  $~5 \pm ~3$ \cr
   (13)    &  $~9 \pm ~3$ \cr
  (16.3)   &  $~8 \pm ~2$ \cr
  (16.5)   &  $~4 \pm 18^*$ \cr
  (16.9)   &  $~8 \pm 18^*$ \cr
  (17.2)   &  $~6 \pm 18^*$ \cr
  (18.1)   &  $~1 \pm ~3$ \cr
  (18.2)   &  $~2 \pm ~3$ \cr
  (18.3)   &  $~8 \pm 18^*$ \cr
  (18.4)   &  $~3 \pm 18^*$ \cr
           &              \cr
\noalign{\hrule}
\noalign{\hrule}
}}$$
*Results which use any data from
Ref.\ [19] are marked with an asterisk. The sum rules
in Eqs.\ (16.5), (16.9), (17.2), (18.3), and (18.4)
cannot be tested solely with the data from Ref.\ [18].

\vfill\eject

TABLE II.~~~ Predicted baryon masses in MeV.  Here
$M_A$ and $M_S$ denote
the two spin-1/2 baryons (with  $\Lambda$-
and $\Sigma$-type symmetry respectively),
$M^*$ denotes spin-3/2 baryons, and $M_B$ denotes baryon
masses spin-averaged according to the
presciption of Ref.\ [15]. The predicted mass values
are determined, wherever possible, using mass
sum rules in which the values of all masses but one
are known from experiment. If more than one such
formula exists for a given hadron, an average of the various
predictions is taken. The results obtained this way are
exploited to obtain new
mass values from sum rules containing
more than one hadron with mass not yet measured.
The errors in the predicted
masses are estimated to be on the order of 20
MeV or less for a baryon containing up to one heavy quark;
on the order of 30 MeV if the baryon contains two $c$
quarks, 40 MeV if one $c$ and one $b$ quarks are present,
and 50 MeV if there are two $b$ quarks.

\vskip .3cm
$$\vbox {\tabskip=0pt \offinterlineskip
\halign {\strut\tabskip=1em plus2em\quad \hfil #\hfil
&\quad\hfil # \hfil &\hfil #\hfil &\hfil #\hfil
& \vrule# & \quad\hfil #\hfil\quad
&\vrule# &\quad\hfil #\hfil\quad &\quad\hfil #\hfil\quad
 &\quad \hfil #\hfil \cr
\noalign{\hrule}
\noalign{\hrule}
       &             &              &               &&
           &&         &         &         \cr
\multispan4 Quark content and symbol                &&
           &&         &         &         \cr
       &             &              &               &&
   $M_B$   &&  $M_A$  &  $M_S$  &  $M^*$  \cr
       &   $M_A$     &     $M_S$    &      $M^*$    &&
           &&         &         &         \cr
       &             &              &               &&
           &&         &         &         \cr
\noalign{\hrule}
       &             &              &               &&
           &&         &         &         \cr
 $qqq$ &             &     $N$      &    $\Delta$   &&
   1086    &&         &~~939$^a$&~1232$^a$\cr
 $qqs$ & $\Lambda$   &   $\Sigma$   &   $\Sigma^*$  &&
   1270    &&~1116$^a$&~1193$^a$&~1385$^a$\cr
 $ssq$ &             &    $\Xi$     &     $\Xi^*$   &&
           &&         &~1318$^a$&~1533$^a$\cr
 $sss$ &             &              &   $\Omega$    &&
           &&         &         &~1672$^a$\cr
 $qqc$ & $\Lambda_c$ & $\Sigma_c$   &  $\Sigma_c^*$ &&
   2450    &&~2285$^a$&~2453$^a$&~2530$^a$\cr
 $qsc$ &  $\Xi_c$    &   $\Xi_c'$   &    $\Xi_c^*$  &&
   2588    &&~2468$^a$&~2582$^b$&~2651$^b$\cr
 $ssc$ &             & $\Omega_c$   &  $\Omega_c^*$ &&
           &&         &~2710$^a$&   2775  \cr
 $qqb$ & $\Lambda_b$ & $\Sigma_b$   &  $\Sigma_b^*$ &&
   5783    &&~5627$^a$&~5818$^a$&~5843$^a$\cr

 $qsb$ &   $\Xi_b$   &   $\Xi_b'$   &    $\Xi_b^*$  &&
           &&         &   5955  &   5984  \cr
 $ssb$ &             & $\Omega_b$   &  $\Omega_b^*$ &&
           &&         &   6075  &   6098  \cr
 $ccq$ &             & $\Xi_{cc}$   &  $\Xi_{cc}^*$ &&
           &&         &   3676  &   3746  \cr
 $ccs$ &             &$\Omega_{cc}$ &$\Omega_{cc}^*$&&
           &&         &   3787  &   3851  \cr
 $qcb$ & $\Xi_{cb}$  & $\Xi_{cb}'$  &  $\Xi_{cb}^*$ &&
   7062    &&  7029   &   7053  &   7083  \cr
 $scb$ &$\Omega_{cb}$&$\Omega_{cb}'$&$\Omega_{cb}^*$&&
   7151    &&  7126   &   7148  &   7165  \cr
 $bbq$ &             & $\Xi_{bb}$   &  $\Xi_{bb}^*$ &&
           &&         &         &  10398~ \cr
 $bbs$ &             &$\Omega_{bb}$ &$\Omega_{bb}^*$&&
           &&         &         &  10483~ \cr
       &             &              &               &&
           &&         &         &         \cr
\noalign{\hrule}
\noalign{\hrule}
}}$$

$^a$ Input mass from experiment [18,19].

$^b$ After this work was completed, we learned that
the $\Xi_c'$ and $\Xi_c^*$ have been observed [24].
Preliminary values of their masses are $2573\pm 4$
and $2643 \pm 4$ respectively, in agreement with our
predictions within our stated errors.
\vfill\eject

\bye